\documentclass[aps,prd,twocolumn,superscriptaddress,nofootinbib,floatfix,
               longbibliography]{revtex4-2}

\usepackage{amsmath,amssymb,bm}
\usepackage{graphicx}
\usepackage{booktabs}
\usepackage{xcolor}
\usepackage{hyperref}
\definecolor{darkblue}{rgb}{0.10,0.20,0.45}
\hypersetup{colorlinks=true,linkcolor=darkblue,citecolor=darkblue,
            urlcolor=darkblue}

\graphicspath{{figs/}{./}}

\newcommand{\mchi}{m_\chi}
\newcommand{\MZB}{M_{Z_B}}
\newcommand{\ZB}{Z_B}
\newcommand{\GN}{G_N}
\newcommand{\vmin}{v_{\min}}
\newcommand{\sigN}{\sigma_N}
\newcommand{\dd}{\mathrm d}
\newcommand{\ie}{\textit{i.e.}}

\begin{document}

\makeatletter
\@namedef{b@apsrev41Control}{}
\makeatother
\nocite{apsrev41Control}

\title{Pseudo-Dirac Inelastic Dark Matter in the Leptophobic $U(1)_B$ Model:\\
Confronting the LUX-ZEPLIN High-Recoil Event with Collider Searches}

\author{Xin-Yu Du}
\email{2020dxy@sjtu.edu.cn}
\author{Wenjie Huang}
\email{weh68@pitt.edu}
\author{Keping Xie}
\email[Corresponding author: ]{kepingxie@sjtu.edu.cn}
\affiliation{State Key Laboratory of Dark Matter Physics \& Key Laboratory for Particle Astrophysics and Cosmology (MOE) \& Shanghai Key Laboratory for Particle Physics and Cosmology, Tsung-Dao Lee Institute \& School of Physics and Astronomy, Shanghai Jiao Tong University, Shanghai 201210, China\looseness=-1}

\date{\today}

\begin{abstract}
Interpreting the $248$~keV nuclear-recoil candidate reported by LUX-ZEPLIN (LZ) as
an endothermic transition needs two ingredients usually put in by hand: a
stable dark state, and a coherent current that is purely off diagonal. Gauging
baryon number supplies both. For the minimal dark charges $B(\chi_L)=-3/2$ and
$B(\chi_R)=-9/2$, the scalars that generate the dark masses leave an exact
residual parity $P=(-1)^{6B}$ stabilizing the lightest state, while the same
Majorana masses make the vector current transition dominated as an identity
rather than a choice. At $\mchi=1$~TeV and $\delta=300$~keV the model is
over-determined: the observed LZ interval, a leading-order relic estimate and
the LHC dijet limit meet at the single perturbatively acceptable point
$(\MZB,g_B)\simeq(1.44~\mathrm{TeV},0.233)$. That point is allowed by currently existing searches but sits just below the trigger-level dijet boundary, so it is
alive and not yet decided---and a re-run of that analysis on the Run-3 data at
$\sqrt s=13.6$~TeV, already on tape, will exclude it or find it. Nor is the
interpretation confined to that splitting: at $\delta=200$~keV, the same
construction predicts a heavier mediator at comparable coupling,
$\MZB\simeq2.4$--$3.0$~TeV with $g_B\simeq0.15$--$0.34$, a range the HL-LHC
dijet program will probe.
\end{abstract}

\maketitle

\section{Introduction}

The LZ collaboration has extended its nuclear-recoil search far above the
classic weakly interacting massive particle (WIMP) window and reports a single
candidate event at $E_{\rm obs}=248\pm23_{\rm stat}\pm23_{\rm sys}$~keV in
$2.84$~tonne-year of exposure, with a low known-background expectation and a
global significance of $2.6\sigma$~\cite{LZ:2026}. For elastic scattering this
is an awkward place for a first event: the spectrum falls steeply, the momentum
transfer $q=\sqrt{2m_AE_R}\simeq250$~MeV lies beyond the first diffraction
minimum of the xenon form factor, and a normalization large enough to produce
one event there would overpopulate the few-keV region where essentially all of
the exposure sits. Inelastic dark matter~\cite{TuckerSmith:2001} inverts that
ordering. For an endothermic transition $\chi_1A\to\chi_2A$ within a
pseudo-Dirac pair split by $\delta$, the minimum halo speed $\vmin(E_R)$ is not monotonic: it
diverges as $E_R\to0$ and passes through a minimum at
$E_R^\star=(\mu_A/m_A)\delta$, which for $\mchi=1$~TeV and $\delta=300$~keV on
xenon is $\simeq265$~keV. The candidate then sits at the kinematically most
accessible energy while the threshold itself removes the low-energy region, so
one high-energy event with an empty low-energy signal region is the generic
expectation rather than a fluctuation.

Recoil-level fits favor heavy dark matter and
$\delta=\mathcal O(300)$~keV~\cite{DiMauro:2026,Su:2026}. Prompt
interpretations include Higgsinos and electroweak multiplets
\cite{Wu:2026,Freese:2026,Fan:2026,Yin:2026,DuWang:2026,Smirnov:2026,
Nomura:2026}, a supersymmetric singlino~\cite{Chattopadhyay:2026}, dark-photon
and Peccei--Quinn constructions~\cite{Yamashita:2026,Visinelli:2026},
absorption and axion-portal alternatives~\cite{Lou:2026,Unwin:2026}, seasonal
scattering~\cite{McCabe:2026}, and a neutrino up-scattering
explanation~\cite{Jeesun:2026}; solar capture constrains electroweak
annihilation channels~\cite{Pospelov:2026}, and the empty high-energy sideband
is a general recoil-spectrum test that applies to the present model as
well~\cite{Rodd:2026}. Most of this work fixes the kinematics and takes the
transition current as given, leaving two structural questions open: what makes
the coherent current transition dominated---an off-diagonal vector coupling is
not a free mixing angle but a statement about the Majorana nature of the mass
eigenstates---and what makes $\chi_1$ stable, which in electroweak completions
is usually supplied by an imposed $\mathbb Z_2$.

Both questions have the same answer once the transition current is required to
descend from a gauge symmetry, and that is our starting point. Gauged baryon
number~\cite{FileviezPerez:2018gwo,FileviezPerez:2019dyo} couples the mediator
to quarks alone, so the nucleon coupling is isoscalar, $f_p=f_n$, and the LZ
interval in the $\mathcal O_1^s$ normalization applies directly---without the
factor $\simeq3$ needed for a neutron-dominated, $Z$-mediated current, a
distinction that matters when a leptophobic $Z'$ is placed alongside a
higgsino~\cite{Wu:2026,Freese:2026}. It also removes the dilepton constraint,
and it is not a simplified model: anomaly cancellation fixes the colorless
spectators, and the scalar sector that breaks $U(1)_B$ is the sector that
generates $\delta$. We build that completion for the minimal charges
$B(\chi_L)=-3/2$, $B(\chi_R)=-9/2$ and confront it, through the four-parameter
set $(\mchi,\delta,\MZB,g_B)$, with the recoil, the thermal history, and the
LHC. Two features are structural rather than assumed: the scalar vacuum
expectation values leave an exact residual parity $P=(-1)^{6B}$ that stabilizes
$\chi_1$ without an imposed symmetry, and the Majorana masses that generate
$\delta$ also force the coherent vector current to be purely off diagonal, with
$\delta\ll\mchi$ technically natural since $\mu_{L,R}\to0$ restores a dark
Dirac number. The recoil recast maps the published interval onto a band in
$(\MZB,g_B)$, with a rate proxy selecting $\MZB/g_B=6.23$~TeV, and a
leading-order relic estimate picks out the single perturbatively acceptable
point $(\MZB,g_B)\simeq(1.44~\mathrm{TeV},0.233)$. Because that point lies
below the invisible threshold, the collider sensitivity is carried by dijets
rather than missing energy---the reverse of the usual simplified-model
expectation---and the candidate lies just under the translated ATLAS boundary.
The relic estimate is not a gauge-complete freeze-out calculation, so this
point is a sharply defined target rather than a demonstrated solution.

This article is organized as follows. Section~\ref{sec:model} defines the model considered in this work. Sec.~\ref{sec:recoil} presents the recoil
calculation, the LZ comparison and the relic estimate. Sec.~\ref{sec:collider}
discusses the collider translations and the induced kinetic mixing. Sec.~\ref{sec:conclusions} summarizes our conclusions.

\section{The leptophobic $U(1)_B$ model}
\label{sec:model}

\subsection{Gauge group, anomalies, and dark charges}

We gauge baryon number, $\mathcal G=\mathcal G_{\rm SM}\times U(1)_B$, with the
Standard Model quarks carrying $B=1/3$ and all leptons and the Higgs doublet
neutral. The mixed anomalies, such as $\mathcal A(SU(2)_L^2\times U(1)_B)$ and
$\mathcal A(U(1)_Y^2\times U(1)_B)$, do not cancel within the Standard Model. Following
Refs.~\cite{FileviezPerez:2018gwo,FileviezPerez:2019dyo} we cancel them with a colorless set of Weyl fermions that is vectorlike under $\mathcal G_{\rm SM}$,
\begin{equation}
\begin{array}{lll}
 \Psi_L\sim(1,2,-\tfrac12,B_1), & \Psi_R\sim(1,2,-\tfrac12,B_2),\\
 \eta_R\sim(1,1,-1,B_1), & \eta_L\sim(1,1,-1,B_2),\\
 \chi_R\sim(1,1,0,B_1), & \chi_L\sim(1,1,0,B_2),
\end{array}
 \label{eq:spectators}
\end{equation}
which is anomaly free provided
\begin{equation}
 B_1-B_2=-3 .
 \label{eq:anomaly}
\end{equation}
The anomaly equation leaves the vector charge
$K\equiv(B_1+B_2)/2$ free. We study the explicit benchmark
\begin{equation}
 B(\chi_L)\equiv B_2=-\tfrac32,\qquad
 B(\chi_R)\equiv B_1=-\tfrac92,
 \label{eq:darkcharges}
\end{equation}
\ie\ $K=-3$. This choice is an input, not a consequence of anomaly
cancellation; other values of $K$ define distinct anomaly-free models. Once it
is fixed, the gauge couplings relevant below follow from $(\MZB,g_B)$.

\subsection{Scalar sector, masses, and the residual parity}

We work throughout with the two left-handed Weyl fields $\chi_L$ and
$\chi_R^c$, which carry $B(\chi_L)=-3/2$ and $B(\chi_R^c)=-B(\chi_R)=+9/2$.
Gauge invariance then dictates the charges of the three scalars that generate
the dark masses,
\begin{equation}
 B(S_B)=-3,\qquad B(\Phi_L)=+3,\qquad B(\Phi_R)=-9,
 \label{eq:scalarcharges}
\end{equation}
through\footnote{As outlined in Refs.~\cite{FileviezPerez:2018gwo,FileviezPerez:2019dyo}, the general Lagrangian of the theory allows other gauge-invariant terms involving $\chi$, namely $y_2\,\chi_R^c\bigl(H\!\cdot\!\Psi_L\bigr)
+y_4\,\chi_L\bigl(\widetilde H\!\cdot\!\Psi_R^c\bigr)
+\mathrm{h.c.}$ One can take small $y_{2,4}$ with $|y_{2,4}|v\ll|M_{\Psi}-M_{\chi}|$, so that the mixing between $\chi$ and the neutral component of $\Psi$ is negligible. This is technically natural because $y_{2,4}\to0$ restores independent fermion parities.}
\begin{equation}
 -\mathcal L_{\chi}=\lambda_\chi S_B\,\chi_L\chi_R^c
 +\frac{y_L}{2}\Phi_L\,\chi_L\chi_L
 +\frac{y_R}{2}\Phi_R\,\chi_R^c\chi_R^c+\mathrm{h.c.}
 \label{eq:darkmass}
\end{equation}
The $S_B$ term gives the Dirac mass and the two $\Phi$ terms provide the Majorana
masses of the left- and right-handed components separately, so their charges
are fixed by $B(S_B)=-B(\chi_L)-B(\chi_R^c)$ and
$B(\Phi_{L})=-2B(\chi_L)$, $B(\Phi_{R})=-2B(\chi_R^c)$. With
$\langle S_B\rangle=v_S/\sqrt2$ and
$\langle\Phi_{L,R}\rangle=v_{L,R}/\sqrt2$, the mass matrix in the basis
$(\chi_L,\chi_R^c)$ is
\begin{equation}
 \begin{aligned}
 \mathcal M_\chi&=\begin{pmatrix}\mu_L&M_D\\M_D&\mu_R\end{pmatrix},
 &M_D&=\frac{\lambda_\chi v_S}{\sqrt2},\\
 \mu_{L,R}&=\frac{y_{L,R}v_{L,R}}{\sqrt2}.
 \end{aligned}
 \label{eq:massmatrix}
\end{equation}
This is a \emph{pseudo-Dirac} structure: a Dirac mass $M_D$ dominates, and the
degeneracy of the Dirac fermion it would otherwise describe is lifted by the
small Majorana entries $\mu_{L,R}$ into a closely spaced pair of Majorana mass
eigenstates. For real $|\mu_{L,R}|\ll M_D$ the eigenvalues and splitting are
\begin{equation}
 m_{1,2}=M_D\mp\frac{\mu_L+\mu_R}{2},
 \qquad
 \delta=|\mu_L+\mu_R|
 +\mathcal O\!\left(\frac{\mu^2}{M_D}\right).
 \label{eq:splitting}
\end{equation}
The limit $\mu_{L,R}\to0$ restores a conserved dark Dirac number and merges the
pair back into a single Dirac state, so a $300$~keV splitting of a TeV fermion
is technically natural: it is controlled by a symmetry-restoring parameter, not
by a cancellation between TeV masses. The pseudo-Dirac spectrum posited in the
original inelastic proposal~\cite{TuckerSmith:2001} is thus an output of the
gauge structure here rather than an input, its two defining features---the
smallness of $\delta$ and the Majorana character of $\chi_{1,2}$---following
from the same three Yukawa terms in Eq.~\eqref{eq:darkmass}.
Numerically, $v_L=v_R=100$~GeV and $y_L=y_R=2.12\times10^{-6}$ give
$\delta=300$~keV.

The same scalar charges leave a residual symmetry. Every vacuum charge in
Eq.~\eqref{eq:scalarcharges} satisfies $6B\in2\mathbb Z$, and so does every
Standard Model field ($6B=2$ for quarks, $0$ for leptons and $H$), whereas the
spectators of Eq.~\eqref{eq:spectators} have $6B=-9$ and $-27$. Gauge breaking
therefore leaves the exact residual parity
\begin{equation}
 P=(-1)^{6B},
 \label{eq:parity}
\end{equation}
under which the entire Standard Model and all three scalars are even and all the new fermions are odd. Since the neutral $\chi_1$ is the lightest odd state in the scenario considered here, it is absolutely stable without any imposed symmetry. This is the structural payoff of gauging $B$ rather than postulating a dark $\mathbb Z_2$.

\subsection{Currents and matching}

The gauge current of the dark sector follows from
Eq.~\eqref{eq:darkcharges} alone,
\begin{equation}
 \mathcal L\supset g_B \ZB^\mu\,
 \overline\chi\gamma_\mu\!\left(K-\tfrac32\gamma^5\right)\!\chi
 =g_B \ZB^\mu\,\overline\chi\gamma_\mu\!
 \left(-3-\tfrac32\gamma^5\right)\!\chi .
 \label{eq:darkcurrent}
\end{equation}
After the Majorana splitting of Eq.~\eqref{eq:massmatrix} the mass eigenstates
$\chi_{1,2}$ are Majorana fermions, for which a diagonal vector bilinear
vanishes identically. Hence, to leading order in $\mu_{L,R}/M_D$,
\begin{equation}
 \overline\chi\gamma^\mu\chi=i\,\overline\chi_2\gamma^\mu\chi_1,
 \qquad
 \overline\chi_i\gamma^\mu\chi_i=0,
 \label{eq:offdiag}
\end{equation}
so the coherent vector current is \emph{exactly} transition dominated with no
small mixing angle imposed by hand, while the axial current stays diagonal and
is velocity suppressed in scattering: the inelasticity is a consequence of the
Majorana masses that generate $\delta$, not an independent assumption.

On the visible side the quark coupling is $g_q=g_B/3$ and the nucleon coupling
is $3g_q=g_B$, so integrating out the mediator at $q^2\ll\MZB^2$ gives an
isoscalar contact interaction with
\begin{equation}
 f_p=f_n\equiv\GN=\frac{|K|\,g_B^2}{\MZB^2}=\frac{3g_B^2}{\MZB^2},
 \qquad
 \sigN=\frac{\mu_N^2}{\pi}\GN^2,
 \label{eq:matching}
\end{equation}
with $\mu_N=\mchi m_N/(\mchi+m_N)$. The mediator mass is itself fixed by the
same vacuum expectation values,
\begin{equation}
 \MZB^2=g_B^2\left(9v_S^2+9v_L^2+81v_R^2\right)\simeq9g_B^2v_S^2,
 \label{eq:mzb}
\end{equation}
since $v_{L,R}\ll v_S$---the $\Phi$ terms amount to $2.4\%$ of $\MZB^2$ at the
benchmark values quoted below---so that $v_S=\MZB/(3g_B)$ and
$\lambda_\chi=\sqrt2\,\mchi/v_S$ are outputs rather than inputs.

The gauge and dark-mass sector relevant to the recoil is parameterized by
$(\mchi,\delta,\MZB,g_B)$ for the charge choice in
Eq.~\eqref{eq:darkcharges}. The complete theory also
contains scalar masses and mixings, spectator masses, and kinetic mixing. We
hold those auxiliary inputs fixed where needed, take $\mchi=1$~TeV and
$\delta=300$~keV, and scan the $(\MZB,g_B)$ plane. Direct detection depends on
$\GN$ in Eq.~\eqref{eq:matching}, while thermal and collider observables can
separate $\MZB$ from $g_B$.

\section{Direct detection and relic density}
\label{sec:recoil}

\subsection{Recoil rate and the LZ normalization}

For isotope $A$ with mass $m_A$, reduced mass $\mu_A$ and Helm form factor
$F_A$~\cite{Helm:1956,LewinSmith:1996}, the spin-independent differential cross
section in the contact limit is
$\dd\sigma_A/\dd E_R=m_AA^2\GN^2F_A^2(E_R)/(2\pi v^2)$. Energy conservation in
$\chi_1+A\to\chi_2+A$ then imposes the standard endothermic threshold
condition, obtained by Tucker-Smith and Weiner when inelastic dark matter was
introduced~\cite{TuckerSmith:2001} and used in the form quoted here
since~\cite{TuckerSmith:2004,Bramante:2016},
\begin{equation}
 \vmin(E_R)=\frac{m_AE_R/\mu_A+\delta}{\sqrt{2m_AE_R}} .
 \label{eq:vmin}
\end{equation}
It reduces for $\delta\to0$ to the elastic expression
$\vmin=\sqrt{m_AE_R/2\mu_A^2}$~\cite{LewinSmith:1996} and has a minimum at
$E_R^\star=(\mu_A/m_A)\delta$. The accepted spectrum is
\begin{equation}
 \frac{\dd N}{\dd E_R}=\mathcal E\,\epsilon_{\rm LZ}(E_R)\,
 \frac{\rho_\chi}{\mchi}\sum_A\frac{\xi_AA^2\GN^2}{2\pi}
 F_A^2(E_R)\,\eta(\vmin^A),
 \label{eq:rate}
\end{equation}
with $\xi_A$ the isotope mass fraction and
$\eta(v_*)=\int_{v>v_*}f_{\rm lab}(\bm v)v^{-1}\dd^3v$ the standard truncated
Maxwellian integral~\cite{Savage:2006qr}.

We adopt the standard halo model (SHM)---an isotropic Maxwell--Boltzmann
distribution of most probable speed $v_0$, truncated at the galactic escape
speed $v_{\rm esc}$ and boosted to the laboratory frame by the mean Earth speed
$\bar v_E$~\cite{LewinSmith:1996,Savage:2006qr}---with the parameters of
Ref.~\cite{DiMauro:2026},
$\rho_\chi=0.30~\mathrm{GeV/cm^{3}}$ and
$(v_0,v_{\rm esc},\bar v_E)=(238,544,250.2)~\mathrm{km/s}$, so that what
follows is an implementation cross-check rather than a re-fit, and normalize
to $N_{\rm sig}=0.9894$ accepted events in the published $2.84$~tonne-year
exposure. Natural xenon isotopes are summed separately, and the LZ efficiency
is digitized from the collaboration's Supplemental Fig.~S2~\cite{LZ:2026}:
approximately $96\%$ between $14$ and $250$~keV, falling through $50\%$ at
$5.4$ and $269.9$~keV. Since $E_R^\star\simeq265$~keV lies on that upper edge,
the efficiency shape is an important normalization uncertainty.

This one-dimensional calculation is not an official likelihood: it reproduces
neither the LZ detector response nor its Neyman construction.

\begin{figure}[t]
    \centering
    \includegraphics[width=0.99\linewidth]{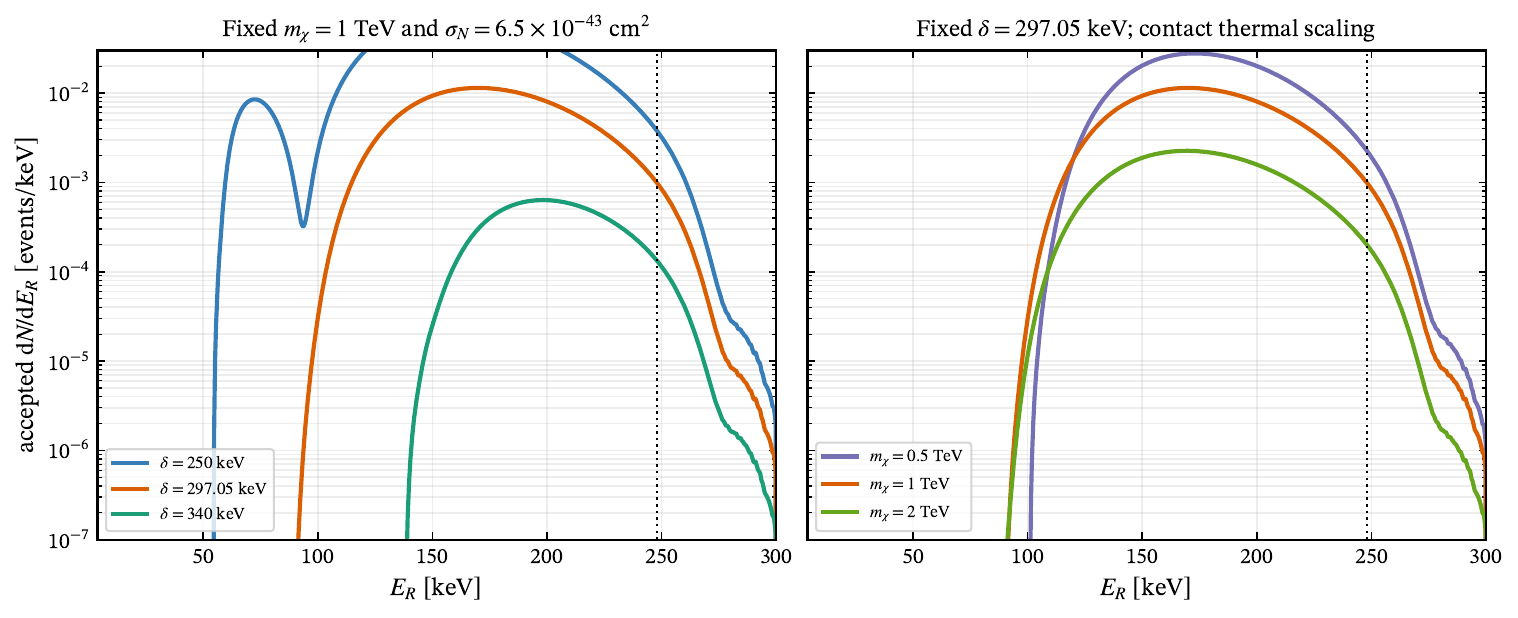}
    \caption{Accepted xenon recoil spectra after the digitized LZ efficiency.
    Left: variation with $\delta$; right: variation with $\mchi$. The vertical
    line marks the $248$~keV candidate, and the dips are Helm-form-factor
    minima.}
    \label{fig:ER}
\end{figure}

At $(\mchi,\delta)=(1~\mathrm{TeV},300~\mathrm{keV})$ the candidate requires
$\vmin\simeq700$--$705~\mathrm{km/s}$, below the adopted cutoff
$v_{\rm esc}+\bar v_E=794~\mathrm{km/s}$. It therefore lies in the halo tail but remains kinematically accessible.

Normalized to $N_{\rm sig}=0.9894$ accepted events, our implementation gives
$\sigN=6.49\times10^{-43}~\mathrm{cm^2}$, \ie
\begin{equation}
 \GN=7.72\times10^{-8}~\mathrm{GeV^{-2}}
 \quad\Longleftrightarrow\quad
 \frac{\MZB}{g_B}=6.23~\mathrm{TeV},
 \label{eq:lzline}
\end{equation}
using Eq.~\eqref{eq:matching}. This normalization is evaluated at the fixed
benchmark $\delta=300$~keV; nearby recoil fits give values around
$295$--$297$~keV~\cite{DiMauro:2026}.

\begin{figure}[t]
    \centering
    \includegraphics[width=0.99\linewidth]{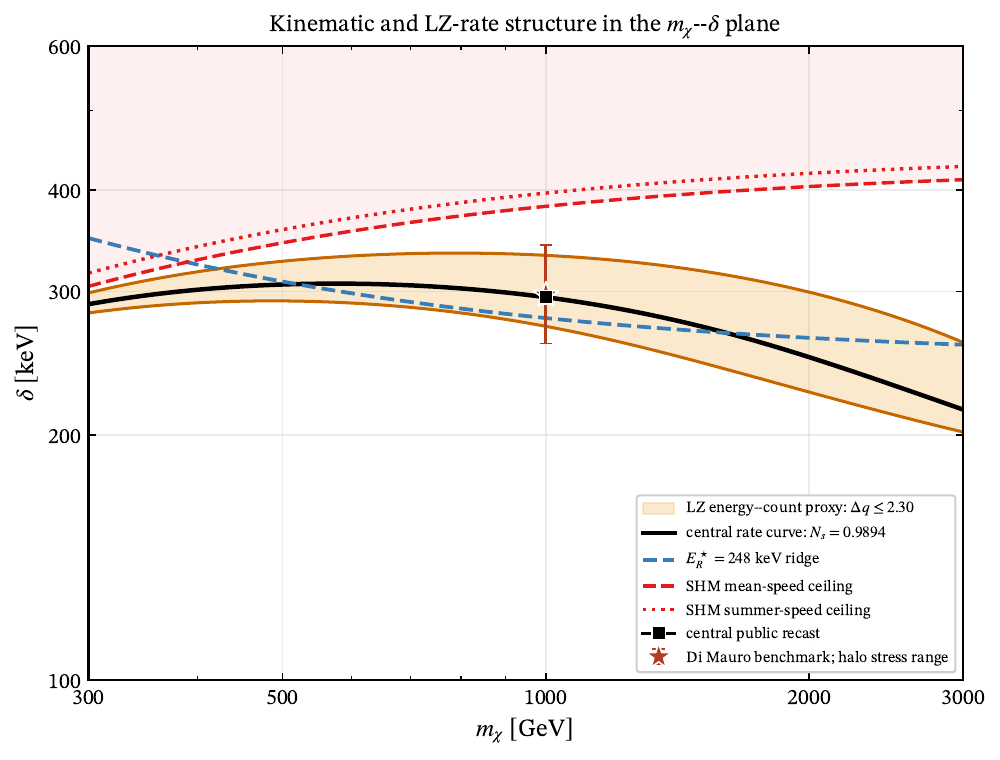}
    \caption{Kinematic structure in the $(\mchi,\delta)$ plane.
The black curve and orange band use the reference normalization of
Eq.~\eqref{eq:proxy-reference}. The black curve selects $N_s=0.9894$;
the orange band is the energy--count proxy region $\Delta q\leq2.30$
defined in Eqs.~\eqref{eq:lz-proxy-likelihood} and
\eqref{eq:lz-proxy-deltaq}, not an official LZ confidence region.
The star shows Ref.~\cite{DiMauro:2026}. Dashed curves show the
$248$~keV ridge and halo ceilings.}
    \label{fig:mchih-delta}
\end{figure}

Figure~\ref{fig:mchih-delta} extends this comparison to the
$(\mchi,\delta)$ plane. For this illustrative scan, both the black curve and
the orange band use the fixed reference contact-target scaling
\begin{equation}
 \sigma_N^{\rm ref}(\mchi)=6.5\times10^{-43}~\mathrm{cm^2}
 \left[\frac{\mu_N(\mchi)}{\mu_N(1~\mathrm{TeV})}\right]^2
 \left(\frac{1~\mathrm{TeV}}{\mchi}\right)^2 .
 \label{eq:proxy-reference}
\end{equation}
The black curve selects $N_s=0.9894$. To display the additional information
carried by the candidate energy, the orange band uses a schematic
single-event energy--count likelihood proxy,
\begin{equation}
 \begin{split}
 \mathcal L_{\rm proxy}(\mchi,\delta,b)\propto{}&
 e^{-(N_s+b)}\big[N_s f_s(E_{\rm obs})+b f_b(E_{\rm obs})\big]\\
 &\times\exp\!\left[-\frac{(b-b_0)^2}{2s_b^2}\right],
 \end{split}
 \label{eq:lz-proxy-likelihood}
\end{equation}
where $E_{\rm obs}=248$~keV and $N_s$ is the accepted yield evaluated with
Eq.~\eqref{eq:proxy-reference}, not a freely fitted signal normalization.
The proxy signal density $f_s$ is obtained by Gaussian-smearing the accepted
recoil spectrum with $s_E=\sqrt{23^2+23^2}\simeq32.5$~keV and normalizing
the resulting shape over $50$--$300$~keV. We take a flat background density
$f_b=1/(250~\mathrm{keV})$ in this window and constrain its yield with
$b_0=0.0106$ and $s_b=0.0008$. These are recoil-level proxy prescriptions,
not the full LZ detector-response or background model.
In particular, $N_s$ retains the accepted true-recoil integral over
$0.25$--$300$~keV, whereas $f_s$ is normalized in the smeared
$50$--$300$~keV window. This mixed selection is retained only as a
diagnostic prescription, not as a selection-consistent extended likelihood.

At each $(\mchi,\delta)$, we maximize over $b\geq0$ and define
\begin{equation}
 \Delta q(\mchi,\delta)=-2\ln
 \frac{\mathcal L_{\rm proxy}(\mchi,\delta,\widehat b)}
 {\displaystyle\max_{\mchi',\delta',b\geq0}
 \mathcal L_{\rm proxy}(\mchi',\delta',b)} ,
 \label{eq:lz-proxy-deltaq}
\end{equation}
with the denominator maximized over $300~\mathrm{GeV}\leq\mchi\leq 3000~ \mathrm{GeV}$
and $150~\mathrm{keV}\leq\delta \leq380~\mathrm{keV}$. The orange band contains points
with $\Delta q\leq2.30$.
At $\mchi=1$~TeV it spans approximately $272$--$332$~keV. The value $2.30$
is the conventional two-parameter likelihood-ratio level corresponding to
about $68.3\%$ coverage in the large-sample Gaussian limit. With only one
event, we use it solely as an illustrative contour level, without assigning
that frequentist coverage. The band need not be centered on the black
fixed-count curve, and is neither an official LZ confidence region nor the
published two-sided $90\%$ interval discussed next.

The two red curves in Fig.~\ref{fig:mchih-delta} are the SHM kinematic
ceilings. Energy conservation requires the reduced-mass kinetic energy to cover
the gap, $\delta\le\mu_Av_{\max}^2/2$, so above them $\vmin(E_R)>v_{\max}$ at
every recoil energy and the rate vanishes identically rather than becoming
small. With $v_{\max}=v_{\rm esc}+\bar v_E$ this gives $\delta_{\max}=383$~keV
on xenon at $\mchi=1$~TeV, rising to $397$~keV in June when Earth's velocity
adds maximally, so the separation of the two curves is the annual modulation of
the threshold. The ceiling reflects the assumed SHM truncation rather than the
model, and below roughly $350$--$400$~GeV it falls through $\delta=300$~keV,
where the interpretation runs out of kinematic room.

\subsection{Comparison with the published LZ interval}

\begin{figure}[t]
    \centering
    \includegraphics[width=0.99\linewidth]{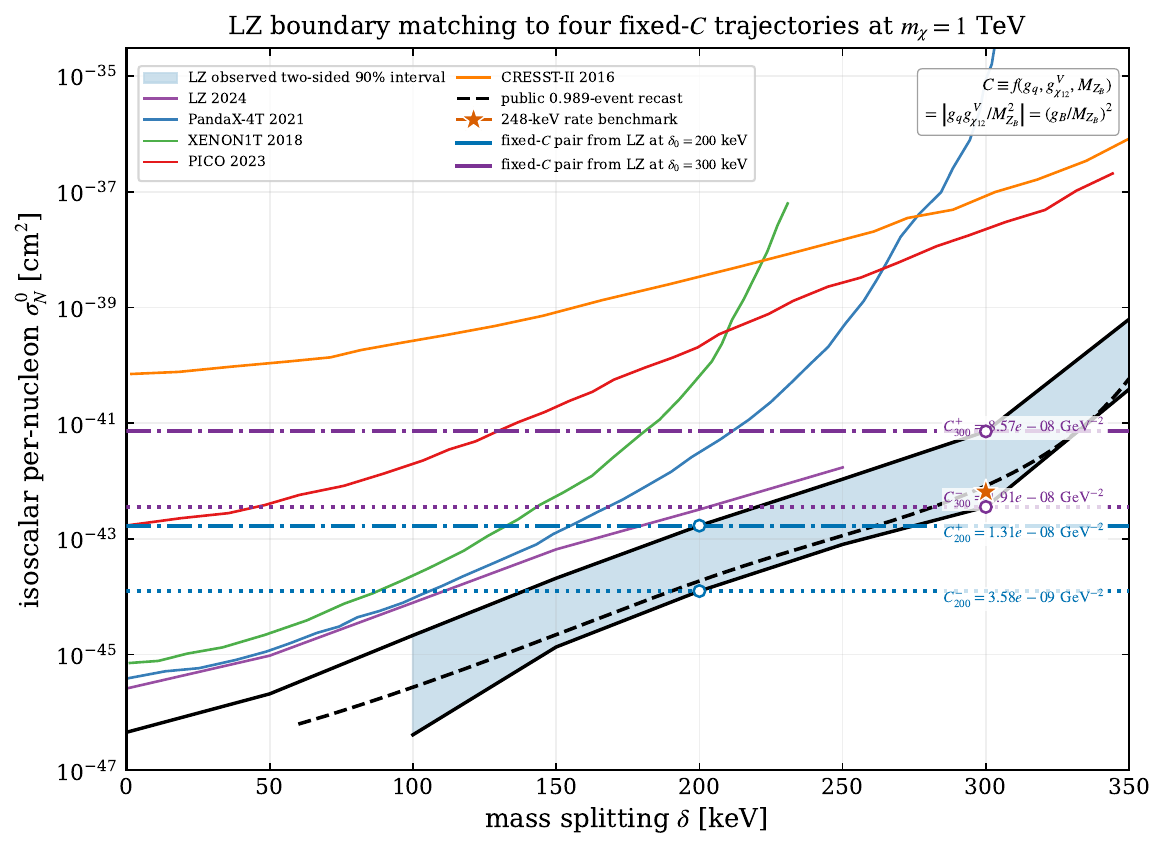}    
    \includegraphics[width=0.99\linewidth]{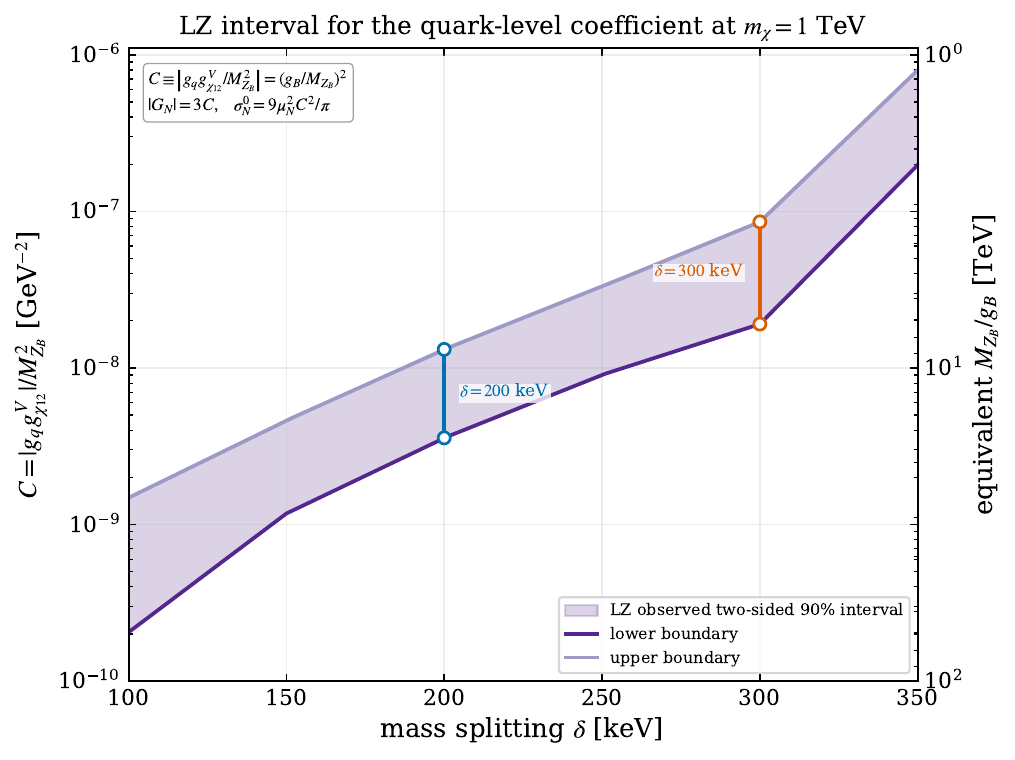}
    \caption{LZ matching at $m_\chi=1$~TeV. Top: the digitized observed
    interval and earlier limits~\cite{LZ:2026}, with boundary values at
    $\delta=200$ and $300$~keV. Bottom: the same interval mapped to
    $C=(g_B/M_{Z_B})^2$; the right axis shows $M_{Z_B}/g_B=C^{-1/2}$. The
    dashed curve is the public rate proxy.}
    \label{fig:SI}
\end{figure}

The interval in Fig.~\ref{fig:SI} is digitized from Supplemental Fig.~S7 of
Ref.~\cite{LZ:2026} in the isoscalar $\mathcal O_1^s$ normalization. Its
comparison with our recoil-energy recast tests normalization and splitting
dependence without reconstructing the experimental likelihood.

For the current charge assignment,
$g_q=g_B/3$ and $g_{\chi_{12}}^V=-3g_B$.
It is useful to distinguish the signed quark-level Wilson coefficient
$\widetilde C$ from its positive magnitude $C$:
\begin{align}
    \widetilde C
    &\equiv \frac{g_qg_{\chi_{12}}^V}{M_{Z_B}^2}
      =-\frac{g_B^2}{M_{Z_B}^2},
      \label{eq:C-signed}\\
    C&\equiv|\widetilde C|
      =\left(\frac{g_B}{M_{Z_B}}\right)^2,
      \qquad G_N=3C,
      \label{eq:C-positive}\\
    \sigma_N^0&=\frac{9\mu_N^2C^2}{\pi}.
      \label{eq:C-sigma}
\end{align}
Here $G_N$ follows the positive-magnitude convention of
Eq.~\eqref{eq:matching}; the signed nucleon Wilson coefficient is
$3\widetilde C$.
Equation~\eqref{eq:C-sigma} is written in natural units.
For the numerical conversion to $\mathrm{cm^2}$ we use
$1~\mathrm{GeV^{-2}}=0.389379\times10^{-27}~\mathrm{cm^2}$.
At fixed $C$, the reference cross section $\sigma_N^0$ is independent
of $\delta$, while the scattering rate retains its splitting dependence
through Eq.~\eqref{eq:vmin}.

Mapping the two observed LZ boundaries separately gives
\begin{align}
 C(200~\mathrm{keV})
 &\in[3.58\times10^{-9},\,1.31\times10^{-8}]
       ~\mathrm{GeV^{-2}}, \label{eq:C-200}\\
 C(300~\mathrm{keV})
 &\in[1.91\times10^{-8},\,8.57\times10^{-8}]
       ~\mathrm{GeV^{-2}}. \label{eq:C-300}
\end{align}
Interpolation is linear in $\delta$ and logarithmic in cross section.

Each interval becomes a band in the mediator plane according to
\begin{equation}
 M_{Z_B}\sqrt{C_{\rm lower}(\delta)}
 \leq g_B \leq
 M_{Z_B}\sqrt{C_{\rm upper}(\delta)}.
 \label{eq:C-plane}
\end{equation}
Figure~\ref{fig:MZBgB} shows the $300$-keV band and a separate $200$-keV
comparison slice. The rate-proxy line is one normalization inside the observed
interval, not an interval boundary.

\subsection{Relic density and the $(\MZB,g_B)$ plane}
\label{sec:plane}

\begin{figure}[t]
    \centering
    \includegraphics[width=0.98\linewidth]{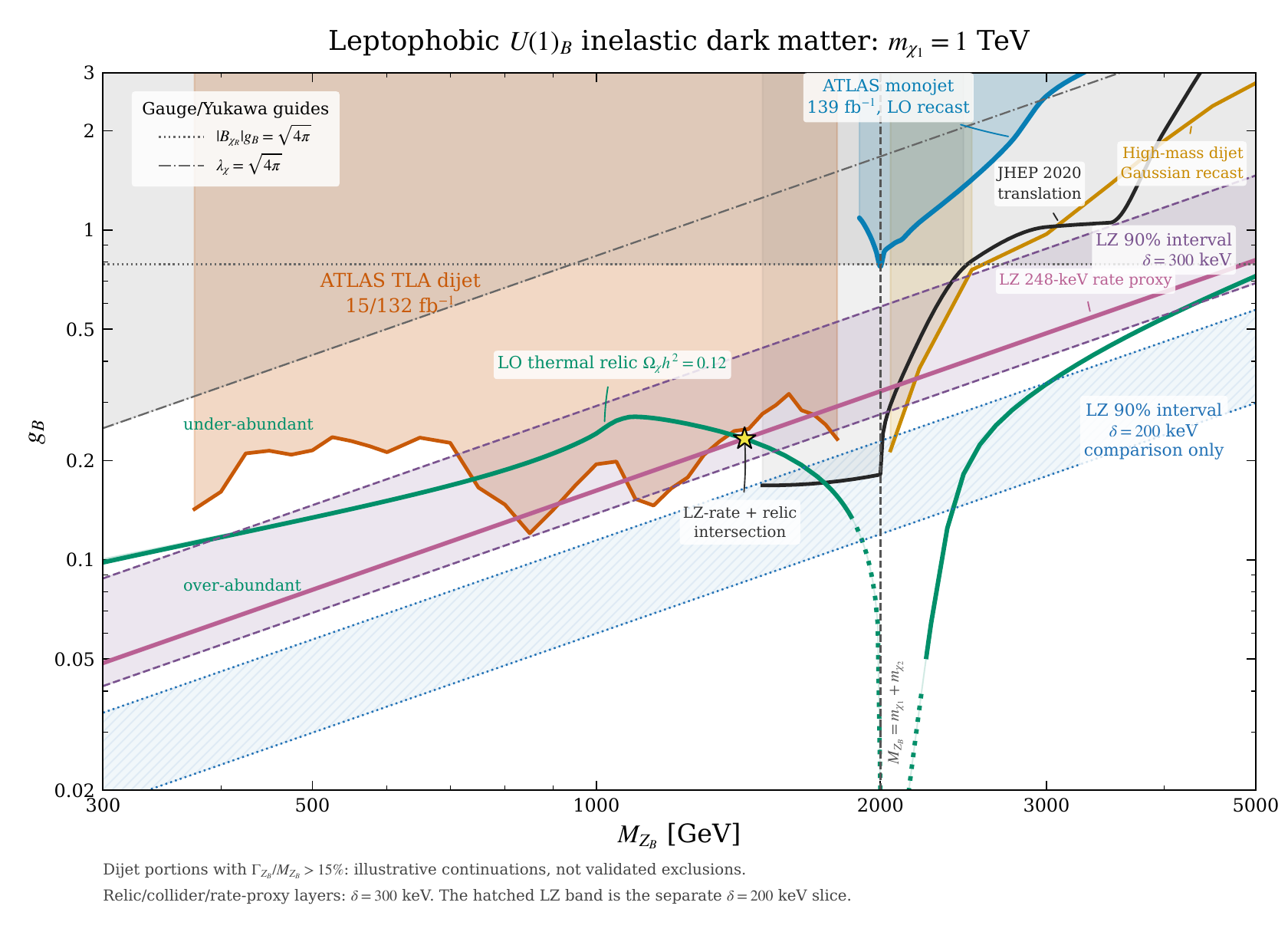}
    \caption{The $(M_{Z_B},g_B)$ plane at $m_{\chi_1}=1$~TeV. LZ bands show
    separate $\delta=300$ and $200$~keV slices; the magenta line is the rate
    proxy. The green curve is a leading-order relic estimate, computed with the
    $U(1)_B$-breaking scalar taken heavy and unmixed with the Standard Model
    Higgs, so that scalar-mediated channels are negligible. Collider
    translations are described in
    Sec.~\ref{sec:collider}~\cite{ATLASTLA:2025,ATLASDijet:2020,
    ATLASMonojet:2021}. LZ bands are two-sided intervals, not exclusions.}
    \label{fig:MZBgB}
\end{figure}

With the charges fixed, the leading coannihilation channel is $\chi_1\chi_2\to
\ZB^*\to q\bar q$; for $\MZB<\mchi$, annihilation to $\ZB\ZB$ is also open. For equilibrium
weights $r_1\simeq r_2\simeq1/2$~\cite{Griest:1990kh},
\begin{align}
 \langle\sigma_{\rm eff}v\rangle_{q\bar q}
 &=\frac{\GN^2\mchi^2}{\pi}\,
 \frac{\MZB^4}{(\MZB^2-4\mchi^2)^2+\MZB^2\Gamma_{\ZB}^2},
 \label{eq:svqq}\\
 \langle\sigma_{\rm eff}v\rangle_{\ZB\ZB}
 &=\frac{(1-r)^{3/2}}{8\pi\mchi^2(2-r)^2}
 \Big[(g_\chi^V)^4+(g_\chi^A)^4 \nonumber\\
 &\qquad\qquad+\tfrac{2(4-3r)}{r}(g_\chi^Vg_\chi^A)^2\Big],
 \label{eq:svvv}
\end{align}
with $r=\MZB^2/\mchi^2<1$, $g_\chi^V=-3g_B$, $g_\chi^A=-3g_B/2$, and the
physical quark thresholds included in $\Gamma_{\ZB}$. Equation~\eqref{eq:svqq}
includes six flavors and the $2r_1r_2=1/2$ coannihilation weight. The $1/r$
term in Eq.~\eqref{eq:svvv} is a longitudinal-vector contribution.

We refer to the relic contour obtained from Eqs.~\eqref{eq:svqq}
and~\eqref{eq:svvv} as the leading-order \emph{estimate}: it combines the
threshold-expansion cross sections with equilibrium coannihilation weights and
the standard freeze-out relation, and omits scalar contribution, thermal averaging,
and any resummation near the pole. It identifies which regions of
$(\MZB,g_B)$ remain in contention rather than determining where the thermal
solution lies. For $g_\chi^A\ne0$, longitudinal-vector production is tied by Ward
identities to the scalar sector that breaks $U(1)_B$; scalar diagrams and full
thermal averaging must be included in a quantitative freeze-out calculation.
Nonperturbative corrections, on the other hand, are not a concern here:
Sommerfeld enhancement and bound-state formation require a mediator light
compared with $\alpha_\chi\mchi$, whereas
$\alpha_\chi=(g_\chi^V)^2/4\pi\simeq0.039$ gives $\alpha_\chi\mchi\simeq39$~GeV,
a factor $\simeq37$ below $\MZB$ at the benchmark.

Figure~\ref{fig:MZBgB} shows the result. In the region $\MZB<\mchi$, where the
$\ZB\ZB$ channel is open, the rate along the LZ line peaks at
$4.0\times10^{-27}~\mathrm{cm^3/s}$ near $\MZB\simeq770$~GeV, a factor
$5.4$ below the canonical value, corresponding to
$\Omega_\chi h^2\simeq0.65$ against the measured
$0.120$~\cite{Planck:2018vyg}. For $\MZB>\mchi$ the $q\bar q$ channel rises
toward the $s$-channel pole at $2\mchi$. The rate-proxy and relic curves
cross twice.

It is worth being precise about what drives the first crossing. Writing
Eq.~\eqref{eq:svqq} as $\langle\sigma_{\rm eff}v\rangle_{q\bar q}
=(\GN^2\mchi^2/\pi)\,P$ with
$P=\MZB^4/[(\MZB^2-4\mchi^2)^2+\MZB^2\Gamma_{\ZB}^2]$, the propagator factor
along the LZ line is $P=0.03$ at $\MZB=770$~GeV, $0.11$ at $1$~TeV, $0.32$ at
$1.2$~TeV and $1.16$ at the crossing. The thermal target is therefore not
reached by resonant enhancement---at the crossing $P$ is within $16\%$ of its
contact-limit value $P\to1$---but by the switching off of the strong
sub-threshold suppression as $\MZB$ approaches $2\mchi$ from below. Since $P$
varies by a factor of a few over a few hundred GeV in this region, the location
of the crossing is correspondingly sensitive to the treatment of the thermal
average and of the width. The second crossing lies beyond the range plotted in
Fig.~\ref{fig:MZBgB}, at $\MZB=7.24$~TeV and $g_B=1.16$.
Treating the fermion and scalar content of Eqs.~\eqref{eq:spectators} and
\eqref{eq:scalarcharges} as active above $\MZB$ and neglecting threshold
corrections, the one-loop coefficient in $\dd g_B/\dd\ln\mu=b_Bg_B^3/16\pi^2$
is $b_B=\tfrac23\sum_{\rm Weyl}B^2+\tfrac13\sum_{\rm scalars}B^2=95.7$, so
that
\begin{equation}
 \Lambda_{\rm LP}=\MZB\exp\!\left[\frac{8\pi^2}{b_Bg_B^2}\right],
 \label{eq:landau}
\end{equation}
which at the second crossing is $13$~TeV, only $1.8\MZB$; we discard that point
as lacking a useful perturbative range. The lower candidate has
$\Lambda_{\rm LP}=5.8\times10^{9}$~GeV and is
\begin{equation}
 \MZB\simeq1.44~\mathrm{TeV},\quad g_B\simeq0.233,\quad
 g_q\simeq0.0777 .
 \label{eq:benchmark}
\end{equation}
The derived quantities are collected in Table~\ref{tab:benchmark}.
Because the point is on the resonance shoulder, Eq.~\eqref{eq:svqq} is not a
precision treatment; a gauge-complete thermally averaged
calculation~\cite{Griest:1990kh} is required to determine whether the
intersection survives and where it lies. For the same reason the relic contour
in Fig.~\ref{fig:MZBgB} is dotted on either side of $\MZB=2\mchi$, where the
threshold expansion fails; only its solid portions are predictions.

The model is not confined to $\delta=300$~keV. Taking the published two-sided
interval at $\delta=200$~keV at face value, the relic contour re-enters the LZ
band \emph{above} the invisible threshold: it lies inside the $200$~keV band
for $\MZB\simeq2.4$--$3.0$~TeV with $g_B\simeq0.15$--$0.34$, against
$g_B=0.233$ at the $\delta=300$~keV benchmark. The smaller splitting is thus
paid for with a heavier mediator at comparable gauge coupling, and hence a
weaker contact interaction: $C$ falls from $2.6\times10^{-8}$ to
$(0.4$--$1.3)\times10^{-8}~\mathrm{GeV^{-2}}$, a factor $2$--$7$. This branch
sits in
a qualitatively different regime, since $\ZB\to\chi_1\chi_2$ is open there and
${\rm BR}(\ZB\to q\bar q)<1$, so dijet and missing-energy searches both apply---%
and both allow it comfortably: the translated high-mass dijet limit runs at
$g_B\simeq0.7$--$1.0$ across that window and the monojet limit at
$g_B\simeq1.3$--$2.5$, a factor of $3$--$5$ above the required coupling, while
the mediator stays narrow, $\Gamma_{\ZB}/\MZB\simeq1.5\%$. The $\delta=200$~keV
solution is therefore \emph{less} constrained by colliders than the $300$~keV
benchmark, at the cost of a heavier mediator, and the two are separated by
whether $\MZB$ lies below or above $2\mchi$.

\begin{table}[t]
 \centering
 \caption{Leading-order benchmark at $\mchi=1$~TeV and
 $\delta=300$~keV. Quantities listed below the mid-table horizontal line are derived. $P$ is the
 propagator factor of Eq.~\eqref{eq:svqq} evaluated at the benchmark. The
 relic density is $0.12$ by construction.}
 \label{tab:benchmark}
 \begin{ruledtabular}
 \begin{tabular}{lc@{\quad}lc}
  Quantity & Value & Quantity & Value \\
  \colrule
  $(B_1,B_2)$ & $(-\tfrac92,-\tfrac32)$ & $K$ & $-3$ \\
  $\MZB$ [TeV] & $1.44$ & $g_B$ & $0.233$ \\
  $g_q$ & $0.0777$ & $\GN$ [GeV$^{-2}$] & $7.9{\times}10^{-8}$ \\
  \colrule
  $C$ [GeV$^{-2}$] & $2.6{\times}10^{-8}$ & $\MZB/g_B$ [TeV] & $6.2$ \\
  $g_\chi^V$ & $-0.699$ & $g_\chi^A$ & $-0.350$ \\
  $\Gamma_{\ZB}/\MZB$ & $2.9{\times}10^{-3}$ & $\MZB/2\mchi$ & $0.720$ \\
  $v_S$ [GeV] & $2060$ & $\lambda_\chi$ & $0.687$ \\
  $y_L=y_R$ & $2.12{\times}10^{-6}$ & $v_L=v_R$ [GeV] & $100$ \\
  $b_B$ & $95.7$ & $\Lambda_{\rm LP}$ [GeV] & $5.8{\times}10^{9}$ \\
  $P$ & $1.16$ & $|\epsilon(\MZB)|$
   & $9.5{\times}10^{-3}$ \\
 \end{tabular}
 \end{ruledtabular}
 \vspace{2pt}

 \raggedright\footnotesize The leading-order estimate admits a second solution at
 $\MZB=7.24$~TeV, discarded because $g_B=1.16$ there places the Landau pole of
 Eq.~\eqref{eq:landau} at only $1.8\,\MZB$.
\end{table}

\section{Collider searches}
\label{sec:collider}

\subsection{Why dijets, and not missing energy}

The benchmark of Eq.~\eqref{eq:benchmark} lies below the invisible threshold,
$\MZB\simeq1.44~\mathrm{TeV}<m_{\chi_1}+m_{\chi_2}=2~\mathrm{TeV}$, so
$\ZB\to\chi_1\chi_2$ is kinematically closed and
${\rm BR}(\ZB\to q\bar q)=1$. This single fact reorganizes the collider
phenomenology, and its consequence is visible directly in
Fig.~\ref{fig:MZBgB}. The monojet
translation~\cite{ATLASMonojet:2021} is a sharp V whose minimum sits on the
$\MZB=m_{\chi_1}+m_{\chi_2}$ line, turning up almost vertically as $\MZB$
falls below threshold and the invisible decay closes. More telling than its
shape is its depth: at that minimum the translated limit is $g_B\simeq0.8$
against $g_B\simeq0.32$ for the LZ line at the same mass. Missing energy thus
falls short of the direct-detection band by a factor $\simeq2.5$ at its single
most sensitive point, and by a larger factor everywhere else. Below threshold the same final state survives only through a far
off-shell mediator with $\hat s>(m_{\chi_1}+m_{\chi_2})^2$; a Born-level
estimate with CT25NNLO parton distributions~\cite{Ablat:2025gdb} gives
$\sigma(pp\to\chi_1\chi_2)\simeq8\times10^{-2}$~fb inclusively, before
requiring initial-state radiation. It is only a production diagnostic; a
mono-$X$ prediction requires showering and the experimental
selection~\cite{CMSMonojet:2021,ATLAS:2020monophoton}. Dijet
production is resonant with unit visible branching fraction and is therefore
the leading collider test at this benchmark.

\subsection{Placing the benchmark against the dijet searches}

Two ATLAS datasets underlie the dijet translations in Fig.~\ref{fig:MZBgB}.
The $132~\mathrm{fb^{-1}}$ trigger-level analysis covers
$375$--$1800$~GeV~\cite{ATLASTLA:2025}; the $139~\mathrm{fb^{-1}}$ high-mass
search starts at $1.1$~TeV~\cite{ATLASDijet:2020}. Because
${\rm BR}(\ZB\to q\bar q)=1$
throughout the region below threshold, and because the mediator is narrow at
the benchmark, $\Gamma_{\ZB}/\MZB=2.9\times10^{-3}$, these limits transfer to the present model without a lineshape correction; the
only model dependence in the translation is the relation $g_q=g_B/3$ between
the quark coupling and the gauge coupling. That transfer holds only while the
mediator stays narrow, which it is not throughout the plane: above the
invisible threshold $\ZB\to\chi_1\chi_2$ opens with $|g_\chi^V|=3g_B$, nine
times the quark coupling, and the width rises steeply. The dijet curves in Fig.~\ref{fig:MZBgB} are drawn with continuous
solid lines for presentation, but their portions with
$\Gamma_{\ZB}/\MZB>15\%$ remain illustrative continuations rather
than validated exclusions, and nothing here rests on them.
The $15\%$ criterion is an operational width warning in this recast,
not an exact identification of a Gaussian template width with the
intrinsic decay width. The benchmark sits a factor $\simeq50$
below this warning threshold.
Figure~\ref{fig:MZBgB} shows two
readings of the high-mass search, a direct translation of the published
coupling limit and a Gaussian-lineshape recast; they agree closely, and
although the search begins at $1.1$~TeV neither translated contour reaches
below $\MZB\simeq1.5$~TeV, so the trigger-level analysis alone tests the
benchmark.

That test is closer than a smooth boundary would suggest. The LZ line and the
trigger-level limit run at similar slopes, and the observed limit is jagged, so
the two cross repeatedly rather than once. Reading along $\MZB/g_B=6.23$~TeV, the observed
limit lies below the LZ line near $\MZB\simeq0.9$, $1.1$--$1.3$ and
$1.7$--$1.8$~TeV, and above it near $0.8$, $1.0$ and $1.4$--$1.6$~TeV. The
benchmark falls in the last of those windows: $g_q\simeq0.078$ against a
translated limit of $\simeq0.084$, below it by under $10\%$. The point survives
not because it is comfortably weakly coupled, but because the observed limit
happens to fluctuate upward there.

The translation itself is cleaner than that margin suggests. The published
bound is quoted on the quark coupling for the same $s$-channel process, so any
common higher-order normalization cancels between the ATLAS limit and our
prediction; and since a spin-1 resonance decaying to massless quarks has the
same $1+\cos^2\theta^*$ distribution for vector and axial couplings, the
acceptance carries over. The benchmark is therefore allowed by currently existing dijet searches---but by a margin smaller than the spread of the observed limit
itself, so present data cannot decide it either way.

What will decide it is exposure, and the required exposure already exists. All
four searches translated in Fig.~\ref{fig:MZBgB} use Run-2 data at
$\sqrt s=13$~TeV, and for a background-limited bump hunt the excluded coupling
scales as $\mathcal L^{-1/4}$. Adding the Run-3 sample at $\sqrt s=13.6$~TeV to
the present $132~\mathrm{fb^{-1}}$ brings the combined dataset to several
hundred $\mathrm{fb^{-1}}$ and moves the boundary at $1.44$~TeV from
$g_q\simeq0.084$ to $\simeq0.064$--$0.068$, with a further few per cent from
the higher collision energy. The benchmark, at $g_q=0.078$, would then sit
$15$--$20\%$ \emph{above} the limit rather than $7\%$ below it. This is a quick
test rather than a programmatic one: the data are already recorded, the
analysis exists, and no new technique or trigger strategy is required---only a
re-run of Ref.~\cite{ATLASTLA:2025} on the $13.6$~TeV sample. The HL-LHC would
settle it more decisively still, its $3000~\mathrm{fb^{-1}}$ placing the point
a factor $\simeq2$ above the limit, a factor $\simeq4$ in cross section, but
the answer need not wait that long. The $\delta=200$~keV branch is a target
for the HL-LHC rather than for Run~3, since it predicts a heavier mediator: the
projected reach of $g_B\simeq0.32$--$0.46$ over $\MZB\simeq2.4$--$3.0$~TeV
comes within some $30\%$ of the required $g_B\simeq0.34$ at the top of that
range, so the HL-LHC dijet programme probes it directly while leaving the
lighter end open. The CMS three-jet
search~\cite{CMSDijet:2020} does not bear on this: its published domain ends at
$700$~GeV.

\subsection{Kinetic mixing}

A leptophobic mediator is not automatically dileptophobic. Hypercharge--baryon
kinetic mixing~\cite{Holdom:1985ag} is an independent renormalized coupling
running as $\dd\epsilon/\dd\ln\mu=g_Yg_Bb_{YB}/16\pi^2$ with
$b_{YB}=\tfrac23\sum_{\rm Weyl}YB$. The chosen charges of
Eq.~\eqref{eq:darkcharges} give $\sum_{\rm Weyl}YB=2-2(B_1+B_2)=14$, so
$b_{YB}=28/3$, and a vanishing boundary value at $10$~TeV yields
$|\epsilon(\MZB)|\simeq9.5\times10^{-3}$. This feeds $\ZB\to\ell^+\ell^-$ at a level
that must be tested at $1.44$~TeV, through a narrow-width calculation against
the ATLAS high-mass dilepton search~\cite{ATLAS:2019erb}; suppressing it to
$|\epsilon(\MZB)|\le10^{-3}$ requires an $89\%$ cancellation of the one-loop
running against an ultraviolet threshold. Because $\epsilon$ is an independent
coupling, this is an added boundary condition rather than a tuning of the gauge
sector.

The recoil calculation also assumes a negligible present-day $\chi_2$
population. Since $\delta<2m_e$ the $e^+e^-$ decay is closed, and depletion
through mixing-induced neutrino decays or inelastic conversion depends on the
same low-energy completion; exothermic $\chi_2$ scattering remains an open
consistency test.

Solar capture is an independent handle, and the model passes it on three counts.
Ten years of IceCube data set world-leading limits on the spin-dependent
WIMP--proton cross section above $\mchi\simeq200$~GeV~\cite{IceCube:2025}, but
these do not apply here: the quark current is purely vector, so no axial quark
coupling---and hence no spin-dependent operator $\mathcal O_4$---is generated,
and $\sigma_{\rm SD}$ vanishes at leading order. Capture on solar hydrogen is
in any case kinematically closed, the inelastic threshold $\delta=300$~keV
exceeding $\mu_{\rm H}v^2/2\simeq10$~keV at the solar core by more than an
order of magnitude; among the abundant species only iron, with
$\delta_{\max}\simeq0.56$~MeV, remains open. Even that cannot yield a neutrino
signal, because $\chi_1\chi_1$ annihilation is $p$-wave and the captured
population is thermalized: $\langle\sigma v\rangle\simeq4\times10^{-35}
~\mathrm{cm^3/s}$ at the core temperature, so capture--annihilation equilibrium
is never approached. This is why the solar bound that constrains electroweak
annihilation channels~\cite{Pospelov:2026} does not bear on this model.

\section{Summary and outlook}
\label{sec:conclusions}

Anomaly-free gauged baryon number supplies both ingredients that an inelastic
reading of the LZ candidate requires, and supplies them from the same source.
Anomaly cancellation against colorless spectators fixes $B_1-B_2=-3$ but leaves
$K=(B_1+B_2)/2$ free; for the minimal choice $K=-3$, gauge invariance forces
the scalars that generate the Dirac and Majorana masses to carry $B=-3$, $+3$
and $-9$, and their vacuum expectation values leave the exact residual parity
$P=(-1)^{6B}$ of Eq.~\eqref{eq:parity}, under which the Standard Model and all
three scalars are even and the new fermions are odd. The same Majorana masses
split the pseudo-Dirac pair into two Majorana states, whose diagonal vector
bilinears vanish
identically, so the coherent current is transition dominated as an identity
rather than as a choice of mixing angle, and the splitting is technically
natural since $\mu_{L,R}\to0$ restores a conserved dark Dirac number. None of
this depends on the numerical benchmark, and it holds for any integer $K$.

At $\mchi=1$~TeV and $\delta=300$~keV the recast maps the published interval
onto $C\in[1.9,8.6]\times10^{-8}~\mathrm{GeV^{-2}}$ and the rate proxy selects
$\MZB/g_B=6.23$~TeV, the comparison being made directly in the isoscalar
$\mathcal O_1^s$ normalization.

The freeze-out estimate crosses that line twice. The upper crossing, at
$\MZB=7.24$~TeV with $g_B=1.16$, places the Landau pole of
Eq.~\eqref{eq:landau} at only $1.8\MZB$ and is discarded; the lower crossing is
Eq.~\eqref{eq:benchmark}. The propagator factor of Eq.~\eqref{eq:svqq} is
$P=0.03$ at $\MZB=770$~GeV and $P=1.16$ there, so the thermal target is reached
as the sub-threshold suppression switches off, not through resonance, and
Sommerfeld enhancement would require $\MZB\lesssim\alpha_\chi\mchi\simeq39$~GeV.
The estimate decouples the scalar sector, to which the longitudinal-vector
amplitude is tied by Ward identities. Since $P$ changes by a factor of a few
near the crossing, it is the \emph{position} of the intersection, not its
existence, that a gauge-complete calculation must settle.

Below the invisible threshold ${\rm BR}(\ZB\to q\bar q)=1$ and the mediator is
narrow, $\Gamma_{\ZB}/\MZB=2.9\times10^{-3}$. Missing-energy searches then have
no resonant channel, and even at their most sensitive point---the threshold
itself---fall short of the LZ line by a factor $\simeq2.5$ in $g_B$, so they do
not constrain the model anywhere in the plane. Dijets do reach it: the
benchmark lies under $10\%$ below the translated trigger-level boundary, and
since higher-order normalization cancels in the transfer and the acceptance
carries over, that margin is real. The point is thus allowed by currently existing
searches and left undecided by them. Run~3 can settle it quickly: the
$13.6$~TeV dataset alone moves the trigger-level boundary below the benchmark,
which would sit $15$--$20\%$ above the limit rather than $7\%$ under it, and it
requires only a re-run of an existing analysis on data already recorded. The
HL-LHC would place the point a factor $\simeq2$ above the limit. The $\delta=200$~keV branch instead predicts a heavier mediator at comparable
coupling, $\MZB\simeq2.4$--$3.0$~TeV with $g_B\simeq0.15$--$0.34$, a range the
HL-LHC dijet programme will probe. Kinetic mixing generates
$|\epsilon(\MZB)|\simeq9.5\times10^{-3}$ from a vanishing boundary value at
$10$~TeV, which must be confronted with the high-mass dilepton search; pushing
it below $10^{-3}$ requires an $89\%$ cancellation against an ultraviolet
threshold.

Independently of the gauge completion, the kinematics is testable now. The
accepted xenon spectrum of Fig.~\ref{fig:ER} peaks at $171$~keV, with $80\%$ of
the rate between $140$ and $218$~keV and essentially none below $100$~keV, so
additional exposure should fill that band rather than the low-energy region.
The threshold $\delta\leq\mu_Av_{\max}^2/2$ also selects targets: at
$v_{\max}=794~\mathrm{km\,s^{-1}}$, a $300$~keV splitting is accessible on
tungsten ($\delta_{\max}=513$~keV), xenon ($382$~keV) and iodine ($371$~keV),
but closed on germanium ($223$~keV) and argon ($126$~keV). A xenon signal with
null results in germanium and argon is thus the expectation rather than a
tension, whereas a confirmed excess in either light target would exclude this
interpretation outright, for any mediator.

The picture is therefore over-determined rather than fitted, and the three
constraints meet where existing LHC data already reach. The immediate tasks are a gauge-complete relic abundance
including the scalar sector, a cosmological history for the depletion of
$\chi_2$, and a showered dijet recast, together with the sideband test of
Ref.~\cite{Rodd:2026} applied to Fig.~\ref{fig:ER}.
Equation~\eqref{eq:benchmark} is a sharply defined target for all of them: if
it survives, a single $248$~keV recoil will have located a leptophobic gauge
boson to within a factor of order unity in both parameters; if it fails, the
failure will be specific and attributable.

\begin{acknowledgments}
This work is supported by the National Natural Science Foundation of China.
All experimental inputs are public: the LZ efficiency and the intervals of
Fig.~\ref{fig:SI} are digitized from Ref.~\cite{LZ:2026}, and the collider
regions of Fig.~\ref{fig:MZBgB} are leading-order translations of
Refs.~\cite{ATLASTLA:2025,ATLASDijet:2020,ATLASMonojet:2021}.
\emph{AI use statement}---The authors used Claude (Anthropic) and ChatGPT (OpenAI) to accelerate parts of the numerical analysis, to develop plotting code, and to draft and edit the manuscript. All AI-generated code was cross-checked against independent implementations, and the authors take full responsibility for the results and the text.
\end{acknowledgments}

\bibliographystyle{apsrev4-2}
\bibliography{reference}
\end{document}